\documentclass[english]{sobraep}

\title{Lens free holographic imaging for urinary tract infection screening}

\author{Gregory N. McKay$^{1}$, Anisha Oommen$^{1}$, Carolina Pacheco$^{1}$, Mason T. Chen$^{1}$, Stuart C. Ray$^{2}$, Ren\'{e} Vidal $^{1}$, Benjamin D. Haeffele$^{1}$, Nicholas J. Durr$^{1}$\\
	\normalsize $^{1}$Department of Biomedical Engineering, Johns Hopkins University, 3400 N. Charles Street, Baltimore, MD, 21218, USA\\
	\normalsize $^{2}$Johns Hopkins University School of Medicine, 733 N. Broadway, Baltimore, MD 21205 USA\\
	\normalsize e-mail: G.N.M.: E-mail: gmckay1@jhmi.edu, N.J.D.: E-mail: ndurr@jhu.edu
}

\begin{document}

\maketitle

\begin{abstract}
	Urinary tract infections (UTIs) are a common condition that can lead to serious complications including kidney injury, altered mental status, sepsis, and death. Laboratory tests such as urinalysis and urine culture are the mainstays of UTI diagnosis, whereby a urine specimen is collected and processed to reveal its cellular and chemical composition. This process requires precise specimen collection, handling infectious human waste, controlled urine storage, and timely transportation to modern laboratory equipment for analysis. Holographic lens free imaging (LFI) can measure large volumes of urine via a simple and compact optical setup, potentially enabling automatic urine analysis at the patient bedside.  We introduce an LFI system capable of resolving important urine clinical biomarkers such as red blood cells, white blood cells, crystals, casts, and E. Coli in urine phantoms. This approach is sensitive to the particulate concentrations relevant for detecting several clinical urine abnormalities such as hematuria, pyuria, and bacteriuria. We show bacteria concentrations across eight orders of magnitude can be estimated by analyzing LFI measurements. LFI measurements of blood cell concentrations are relatively insensitive to changes in bacteria concentrations of over seven orders of magnitude. Lastly, LFI reveals clear differences between UTI-positive and UTI-negative urine from human patients. Together, these results show promise for LFI as a tool for urine screening, potentially offering early, point-of-care detection of UTI and other pathological processes.  
\end{abstract}

\begin{keywords}
	Lens free imaging, holography, urinalysis, urinary tract infection, point of care diagnostics.
\end{keywords}






\section{INTRODUCTION}

Across healthcare, urinary tract infections (UTIs) are a major cause of morbidity, mortality, and economic cost \cite{Gupta2017,Medina2019,Saint2000}. In out-patient settings, UTIs are the most common infection, with approximately 1 in 2 adult women experiencing a UTI in their lifetime \cite{Medina2019}. For in-patient settings, where approximately 25$\%$ of patients have an in-dwelling urinary catheter during their stay, bacteriuria develops in approximately 26$\%$ of catheterized patients \cite{Saint2000}. These catheter-associated UTIs are the most common nosocomial infection in hospital systems \cite{Stamm1991}. Urinary tract infections are often accompanied by hematuria \cite{Sharp2013} and pyuria \cite{Tambyah2000a}, the presence of red blood cells (RBCs) and white blood cells (WBCs) in the urine, respectively. When diagnosed quickly by urinalysis and culture, UTIs can be managed effectively.

Urinalysis is considered the first ever clinical laboratory test \cite{Rai2010}. Over millennia it has been developed into a powerful diagnostic tool. Conventional urinalysis uses a combination of dipstick testing, urine culture, centrifugation, and microscopic evaluation of urine sediment. The detection and quantification of biomarkers such as blood cells, crystals, and casts can indicate specific pathology such as hemorrhage, infection, and kidney injury \cite{Simerville2005,Wu2010}. While many details of genitourinary pathology have been discovered and precisely correlated to urinalysis findings, the process of collecting, storing, and transporting urine remains relatively rudimentary. Urine specimen collection is still often done by hand, requiring techniques such as mid-stream collection and human handling of potentially infectious human waste \cite{Larocco2015}. Further, urine is inherently unstable, requiring low-light refrigeration to prevent the degradation of cells, crystals, and casts \cite{Rai2010,Simerville2005}. This process can yield new contamination or continued, unpredictable bacterial growth after specimen collection, making transportation to the core laboratory and specimen processing time-sensitive to maintain precision of the results \cite{Wu2010}.

Holographic lens free imaging (LFI) is a compact, simple form of digital holography that has many medical and biological applications \cite{Xu2001,Seo2009,Su2012a}. Through the use of a partially coherent light source, a weakly-scattering specimen is illuminated in transmission mode, and a hologram that encodes both phase and amplitude information is recorded with a 2D sensor \cite{Pedrini2002,Repetto2004,Garcia-Sucerquia2006}. A mathematical model of diffraction can then be applied to the hologram to digitally reconstruct the original sample in 3D space \cite{Kreis1998a,Seo2009,Kim2009}. LFI circumvents the conventional tradeoffs of  resolution and depth-of-field that constrain lens-based imaging, and enables the reconstruction of large volumes of sample at high resolution from a single hologram as long as the sample is sparse\cite{Xu2001,Isikman2010,Seo2009}. Thus, LFI is well-suited for the task of  point-of-care urinalysis screening, as large volumes of urine can be probed efficiently with a compact, low-cost device without typical laboratory processing such as centrifugation.

We present an LFI system capable of reconstructing images of many clinically important urine particulates such as RBCs, WBCs, crystals, and casts. We further demonstrate its ability to reconstruct and estimate the concentration of E. Coli, the most common cause of UTI \cite{Gupta2017,Medina2019}, over physiologically relevant ranges. Finally, we show clear differences in holograms obtained from human urine with a positive UTI diagnosis and control human urine without UTI. These results provide the foundation for further development of a bedside LFI system for point-of-care urine screening. We envision that this technology could be implemented directly in-line of a catheter drainage tube, allowing urine to be probed during micturition, and warning clinical staff to trending changes in urine composition in real-time. Further, we believe this technology could be readily adapted to enable low-cost, point of care urinalysis screening in low-resource settings.

\section{Methods}
\subsection{Optical System}
\begin{figure}[h]
\centerline{\includegraphics[width=\columnwidth]{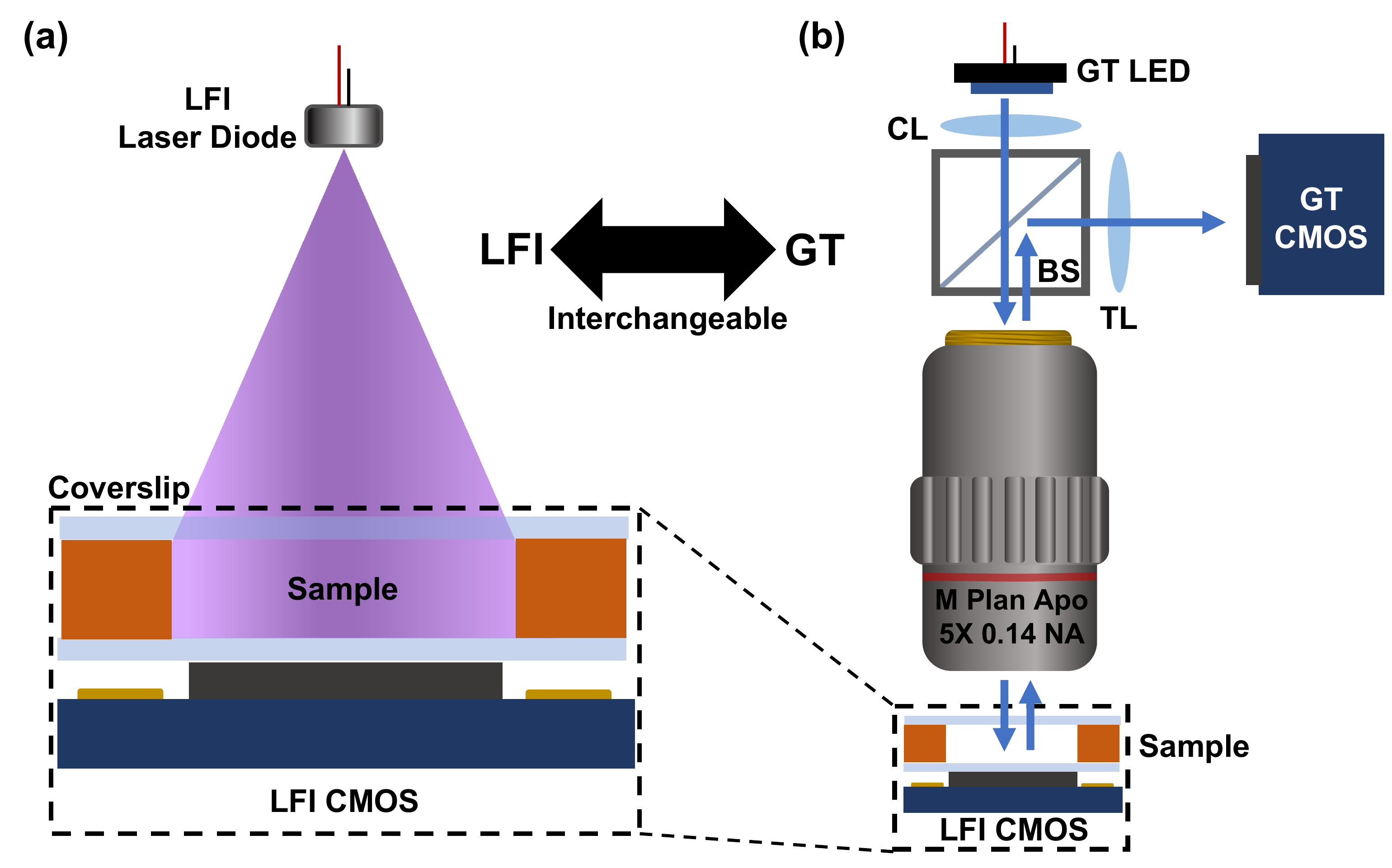}}
\caption{(a) The lens free imaging (LFI) system consists of a 405nm laser diode illuminating the sample placed above a CMOS monochrome board camera. The LFI illumination can be interchanged with a ground truth (GT) microscope (b), through the use of rotation turret. The epi-mode GT microscope consists of an LED, a condensing lens system (CL), a beamsplitter (BS), a tube lens (TL), and a CMOS sensor.}
\label{fig:McKay_LFI_Figure_001}
\end{figure}

Fig. \ref{fig:McKay_LFI_Figure_001} provides an overview of the imaging instrumentation used in this study, which consists of the LFI system built in parallel to a conventional ground truth (GT) microscope. The LFI system consists of a 405 nm laser diode (Thorlabs, L405P20) placed 8 cm away from a 12-megapixel CMOS monochrome board camera (The Imaging Source, DMM 37UX226-ML, 1.85 $\mu$m pixel pitch). A short, visible wavelength of 405 nm was chosen as a wavelength smaller than the short axis of E. Coli (typically 0.5 x 1-2 $\mu$m \cite{Riley1999}) while still being detectable with high quantum efficiency from a silicon sensor.  The LFI sensor provides a 7.40 mm x 5.55 mm field-of-view, and given a sample of 2 mm height, enables the acquisition of holograms over a sample volume of approximately 82 $\mu$L for every hologram reconstruction. The conventional, ground truth (GT), lens-based microscope was built in epi-mode to enable sequential, paired imaging of the same particles between the two systems (Fig. \ref{fig:McKay_LFI_Figure_001}). The GT microscope consists of a 465nm LED (Thorlabs LED465E, LEDMT1E) whose light is collected through a condensing lens illuminator module (Thorlabs WFA2001), and reflected through the 5X 0.14 NA Mitutoyo microscope objective by a 50:50 beamsplitter (Thorlabs BSW10R). Back-scattered light is collected by the same objective, and imaged onto the GT CMOS (Edmund Optics EO-5012M) by an \textit{f} = 200 mm tube lens (TL). The GT microscope is mounted on a z-axis motorized module (Thorlabs ZFM2020 and MCM3001) to enable z-stack acquisition. The LFI laser diode and the GT microscope objective are both mounted on a rotating turret (Thorlabs CSN510 nosepiece) to enable coaxial interchanging between the two imaging systems without moving the sample. Custom MATLAB software is used to acquire raw images. Together, this setup enables the acquisition of paired conventional lens-based images with lens-free holographic reconstructions (Fig. \ref{fig:McKay_LFI_Figure_002}).

\subsection{Sample Preparation}

\subsubsection{Paired LFI and GT Imaging}

For all LFI and GT imaging data presented here, samples were prepared in a 2 mm deep silicone isolator well (Grace Bio Labs, CWS-13R-2.0) placed between two 22 mm x 22 mm square $\#$ 1 coverslips (Tedpella, 260341). A urinalysis control was used to model pathologic human urine (Bio-Rad Liquichek Urinalysis Control Level 2 $\#$437). This is a commercially available, assayed liquid urine phantom used to calibrate clinical urine dipstick and microscopic tests. It primarily contains red and white blood cells, as well as intermittent crystals and casts. To obtain GT images paired to LFI reconstructions, the Bio-Rad urinalysis control was diluted 10X into 1$\%$ low-melting point agar in phosphate buffered saline (PBS). The use of agar for these experiments enabled particles in the urine phantom to be stationary with time, allowing the acquisition of a hologram with the LFI system and paired axial z-stack acquisition with the GT microscope. LFI holograms are reconstructed using \ref{eq_01} and clinically relevant urine particles such as red blood cells, white blood cells, crystals, and casts are shown visually at axial depths throughout the sample.

\subsubsection{RBC and WBC Concentration}

Next, the Bio-Rad urinalysis control was diluted in PBS into six discrete concentrations to generate a physiologically relevant range of red blood cell (0-600 RBCs/$\mu$L) and white blood cell (0-130 WBCs/$\mu$L) concentrations. Of note, a threshold of 8 RBCs/$\mu$L and 8 WBCs/$\mu$L in un-centrifuged urine are typical thresholds used for microscopic diagnosis of hematuria and pyuria, respectively, which are often present in urinary tract infections \cite{Stamm1983,Feld1997,Wang2019,Hannemann-Pohl1999}. This experiment was repeated independently and from the beginning five times, and the number of red and white blood cells was counted in the 3D reconstruction volume manually. The mean and standard deviations of blood cell counts of these five samples was determined and divided by the sample volume over which the cells were counted to estimate their concentrations. Separately, the RBC and WBC concentrations in the Bio-Rad urinalysis control were measured in a hemacytometer (Bright-Line, Z359629) using trypan blue. Due to the large range of blood cell concentrations measured, data is reported with a log-log plot. One data point for WBC concentration measurement returned 0 cells/mL, and is omitted due to its undefined logarithm. Linear correlation is quantified with $R^2$ values, which are applied to all data measurements, including the one WBC measurement omitted from the plot.

\subsubsection{E. Coli Concentration}

E. Coli were added to PBS to assess the capability of the LFI system to resolve bacteria and estimate their concentration in solution. E. Coli (ATCC, 39936) were first incubated at 37$^{\circ}$ C on Tryptic Soy Agar plates with 100 $\mu$g/mL ampicillin (Teknova, 200066-580). A single colony was selected and incubated for 24 hours at 37$^{\circ}$ C in LB Broth with 100 $\mu$g/mL ampicillin. From this stock solution, E. Coli were pelleted using centrifugation (1g x 10 min) and resuspended in PBS. The concentration of E. Coli was measured using a 20 $\mu$m tall Petroff-Hausser bacterial cell counter (Hausser Scientific 3298S22). The stock E. Coli solution in PBS was diluted to span a physiologically relevant range of concentrations (0, 10$^{3}$, 10$^{4}$, 10$^{5}$, 10$^{6}$, 10$^{7}$, and 10$^{8}$ cells/mL), including the typical threshold for UTI in asymptomatic patients (10$^{5}$ cells/mL \cite{Stamm1983}). Again, a manual counting of reconstructed particles was conducted to estimate the E. Coli concentrations. E. Coli could not be individually resolved above 10$^{5}$ cells/mL, so we also applied a textural analysis based upon the gray-level co-occurrence matrix (GLCM) to the raw holograms \cite{Haralick1973}. This was implemented using pre-defined MATLAB functions (graycomatrix and graycoprops) using 25 intensity levels and a one-pixel lateral offset. This  experiment was also repeated five times and the mean and the standard deviations are reported.

\subsubsection{Measuring cross-talk between E. Coli, RBC, and WBC concentrations}

We tested mixed samples to assess the influence of cross talk between concentrations of different components. Using a similar protocol of bacterial selection, growth, and centrifugation, E. Coli were pelleted and then resuspended in 20X dilute Bio-Rad urinalysis control, a concentration chosen to have WBCs present at approximately the threshold of pyuria. E. Coli concentration was again varied across 0, 10$^{3}$, 10$^{4}$, 10$^{5}$, 10$^{6}$, 10$^{7}$, and 10$^{8}$ cells/mL, and the concentration of RBCs and WBCs was measured through manual counting of particles in the reconstruction. E. Coli concentration was again estimated using GLCM texture analysis. This experiment was replicated five times and mean and standard deviations of concentrations are reported. 

\subsubsection{Human UTI Sample Imaging}

Finally, through an IRB-approved protocol (JHU IRB00283348), we acquired discarded urine from six patients at the Johns Hopkins Hospital with paired ground truth diagnosis of positive urinary tract infection (UTI(+), 3 samples) and negative urinary tract infection (UTI(-), 3 samples) controls. These samples were imaged in the LFI system. Because these samples were in solution, they were not imaged with the GT microscope. GLCM textural analysis was applied to five, 1 mm x 1 mm regions of interest (ROIs) for each of the six samples, and mean and standard deviations of these results are reported.

\subsection{Image Acquisition, Processing, and Reconstruction}

All holograms were acquired using a 20 ms exposure and gain setting of 0 dB with 0.5 mW of illumination power incident on the sample window. Single holograms were used for reconstructing paired imaging data between the LFI and GT systems, red and white blood cell concentration estimation, and patient urine samples. E. Coli are often similar in size to the wavelength of light used, and scatter relatively weakly as compared to other larger particles such as RBCs, WBCs, and debris. For phantom data where E. Coli were present in solution and for human urine imaging, videos of 100 frames were acquired at 5 Hz. The time-average image was computed from the 100-frame stack, and subtracted from a single frame to remove signal from stationary particles that were present on the coverglass such as adhered cells, glass imperfections, dust, streaks, and debris.

To reconstruct images from LFI holograms, we implemented a 3D sparse phase recovery reconstruction algorithm developed previously by our group \cite{Haeffele2017,Haeffele2020}. Briefly, sparse regularization is applied to a wide angular spectrum model of diffraction where alternating minimization allows for closed-form updates and recovery of missing phase information in a 3D volume. This reconstruction approach is defined by \ref{eq_01}.

\begin{equation}
\begin{split}
\min_{\{X_j\}^{D}_{j=1},W,\mu} \frac{1}{2} \Arrowvert H \odot W - \mu 1 - \sum_{j=1}^{D} T(z[j]) \ast X_{j} \Arrowvert ^{2}_{F} \\
+ \lambda \sum_{j=1}^{D}\Arrowvert X_{j} \Arrowvert_{1}  \quad  s.t. \vert W \vert = 1.
\end{split}
\label{eq_01}
\end{equation}

Where $H$ is the hologram recorded by the image sensor, $X_{j}$ is the corresponding image at specified depth $z[j]$, $W$ is the estimated phase, $\mu$ is the non-zero background modeling planar illumination, $T(z)$ is the diffraction transfer function according to the wide angular spectrum model, and $\lambda$ is the sparsity parameter. For further description see \cite{Haeffele2017}, including Algorithm 2 in Supplementary Material for pseudocode. We set our axial resolution to 10 $\mu$m, over a 2 mm reconstruction depth, utilized 20 iterations, and a sparsity parameter $\lambda = $ 1.3. For reconstructions shown in Figures \ref{fig:McKay_LFI_Figure_003}, \ref{fig:McKay_LFI_Figure_004}, and \ref{fig:McKay_LFI_Figure_005}, a summed intensity projection over the full 2 mm depth is calculated to visualize the total number of particles in the 3D volume in a single 2D image.

\section{Results and Discussion}

\subsection{Paired LFI and GT Imaging} \label{Paired_LFI_GT_Results}

\begin{figure}[h!]
\centerline{\includegraphics[width=\columnwidth]{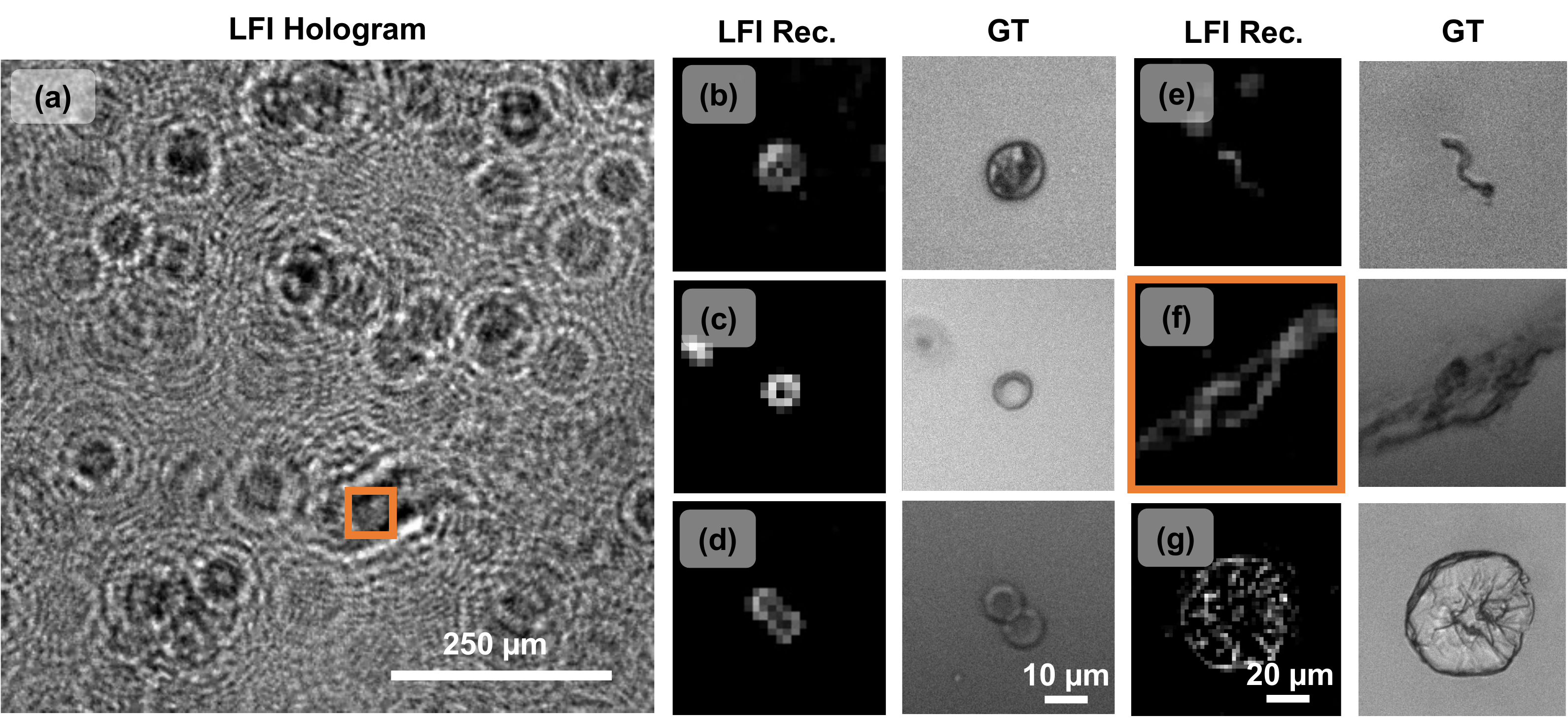}}
\caption{(a) Raw hologram of urinalysis control in low-melting point agar acquired with the LFI system. Note this represents only 1 $\mu$L of the total 82 $\mu$L acquired by the full field of view. A single region of interest (ROI) in the hologram (orange box) is highlighted along with its corresponding reconstruction, (f).  (b)-(g) Paired ROIs of LFI reconstruction (left) and GT microscope (right). (b) A white blood cell, (c)-(d) red blood cells, (e) a fiber, (f) a cast, and (g) a crystal. Note that the 10 $\mu$m scale bar applies to all LFI reconstruction and GT ROIs except the crystal, (g).}
\label{fig:McKay_LFI_Figure_002}
\end{figure}

We first tested the reconstruction quality of clinically important urinary particulates with LFI. To do so, we prepared a stationary urine phantom in low melting point agar that enabled the sequential acquisition of data with the LFI system and with the lens-based ground truth (GT) microscope. Single LFI holograms and a series of z-stack images were obtained with the GT microscope of the same sample. A portion of a raw hologram from the LFI system is shown in Fig. \ref{fig:McKay_LFI_Figure_002}(a). Next, we applied the 3D sparse phase recovery reconstruction algorithm to the hologram, and show regions of interest of reconstructed particles in Fig. \ref{fig:McKay_LFI_Figure_002}(b)-(g) (left), with the paired image of the same particle from the GT microscope (right). These results demonstrate that the LFI system can resolve particles such as blood cells and crystals in a model of human urine. There is a clear difference between RBCs, with their biconcave disk shape, and larger, circular WBCs. Note that while a full 7.40 mm x 5.55 mm hologram is acquired, we show only a small portion of this data in Fig. \ref{fig:McKay_LFI_Figure_002}(a) in order to visualize the fine hologram features. The reconstructed volume in this cropped region is approximately 1 $\mu$L of the 82 $\mu$L sampled in our raw image.

\subsection{RBC and WBC Concentration} \label{RBC_WBC_BioRad_Results}

\begin{figure*}[h!]
\centerline{\includegraphics[width=\linewidth]{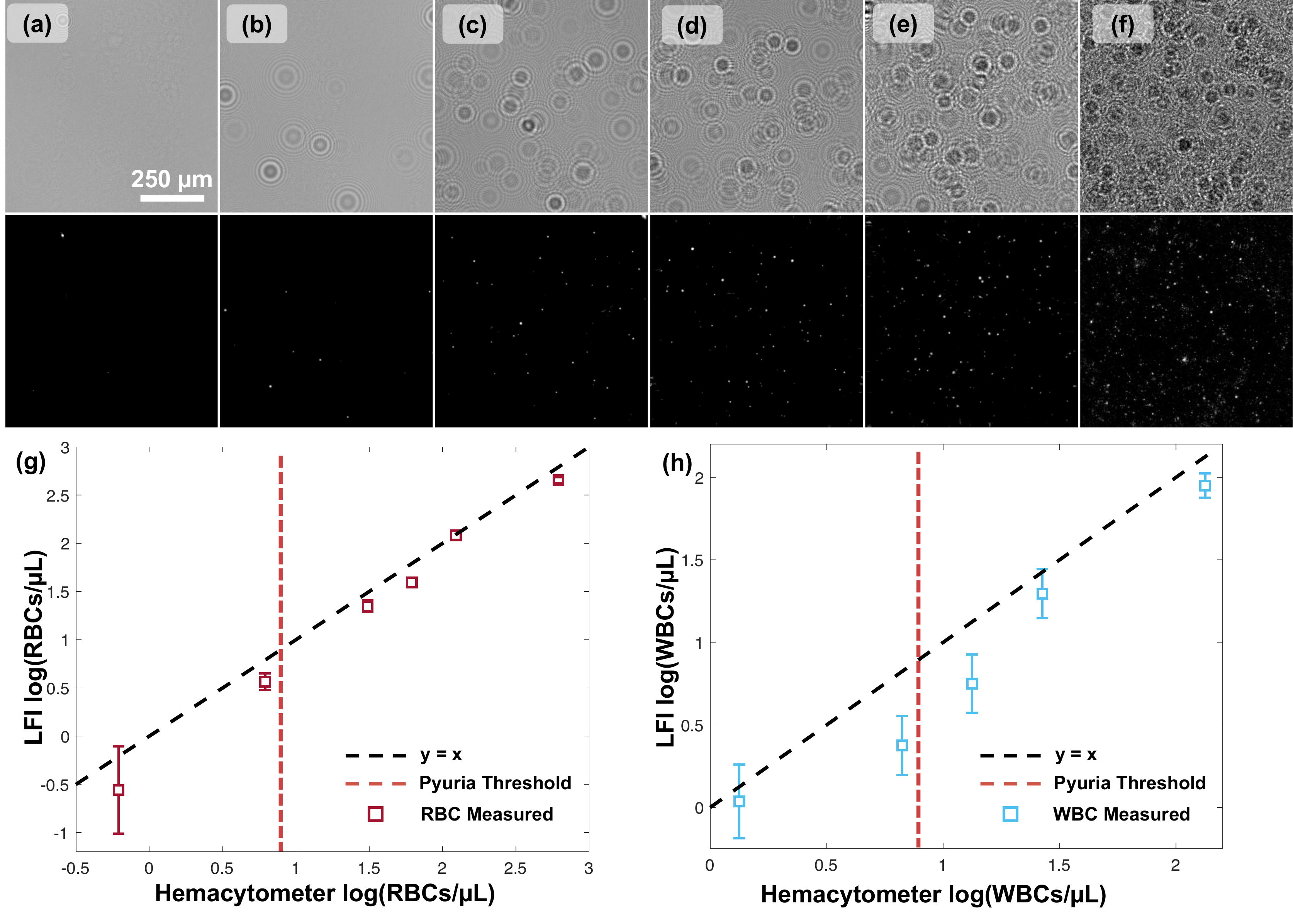}}
\caption{Red and white blood cell concentration estimation using LFI. (a)-(f) Raw hologram (top) and corresponding 2D summed intensity projection of 3D reconstructed volume (bottom) with increasing concentration of RBCs and WBCs. Concentration estimation of (g) red blood cells and (h) white blood cells using LFI vs. a Hemacytometer. The dashed lines represent a linear relationship (y = x), while measured mean and standard deviation of the concentrations over five replicates are shown.}
\label{fig:McKay_LFI_Figure_003}
\end{figure*}

\textcolor{black}{We next tested the accuracy of LFI in measuring concentrations of RBCs and WBCs in a urine solution. To do so, we diluted and imaged the Bio-Rad urinalysis control in PBS and compared the results of the LFI measurement to the concentrations measured with a hemacytometer. Cropped holograms showing increasing RBC and WBC concentration are shown in Fig. \ref{fig:McKay_LFI_Figure_003}(a)-(f) with the corresponding reconstruction shown below. Note the reconstructions are displayed as a summed intensity projection across the full 2 mm sample reconstruction height to enable viewing the particles at different axial locations in the field of view in a single 2D image. The red and white blood cells were counted manually and divided by the reconstructed sample volume to provide an estimation of the RBC and WBC concentrations. The mean and standard deviations from conducting this experiment five times are presented in Fig. \ref{fig:McKay_LFI_Figure_003}(g)-(h)).}

\begin{figure*}[h!]
\centerline{\includegraphics[width=\linewidth]{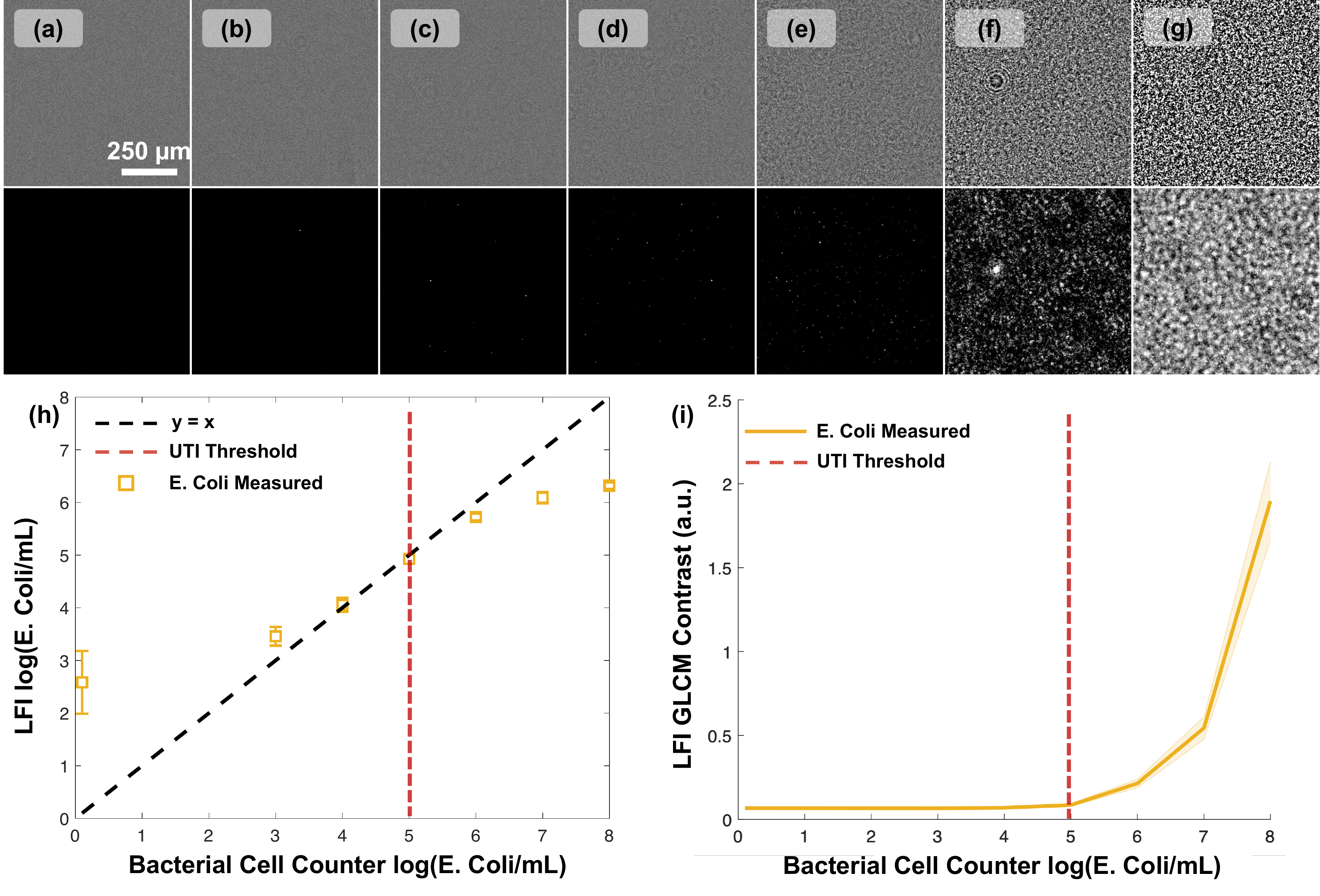}}
\caption{E. Coli concentration estimation in PBS. (a)-(g) Holograms (top) and corresponding summed intensity projections of reconstructions (bottom) of increasing concentration of E. Coli (0, 10$^{3}$, 10$^{4}$, 10$^{5}$, 10$^{6}$, 10$^{7}$, and 10$^{8}$ cells/mL, respectively). The intensity of the reconstructed images is displayed on a log scale to increase visibility at low concentrations. Each field-of-view shows approximately 1 $\mu$L of sample. (h) Manual counting of cells yields concentration estimations (yellow data points) that are accurate from 10$^{3}$ cell/mL up to the UTI threshold (10$^{5}$ cells/mL, red dashed line), however under-estimates E. Coli concentration above the UTI threshold. (i) Gray-level co-occurrence matrix contrast vs. E. Coli concentration shows a positive correlation above the UTI threshold.}
\label{fig:McKay_LFI_Figure_004}
\end{figure*}

\textcolor{black}{From these data, a few important findings emerge. First, we observe a strong linear correlation between the concentration estimation in the LFI system and from the hemacytometer with $R^{2} = 0.9941$ and $R^{2} = 0.9973$ for the RBC and WBC concentrations, respectively. However, despite the strong linear correlation in the data, we observe that for both RBC and WBC concentration estimation, the LFI system under-estimates the cellular concentration, a trend which becomes slightly more pronounced as concentration increases (note the log scale). At the highest concentration imaged, the holograms from individual particles (Fig. \ref{fig:McKay_LFI_Figure_003}(f), top) become quite crowded and are no longer individually separated. At this concentration, the reconstructions (Fig. \ref{fig:McKay_LFI_Figure_003}(f), bottom) become noisy. This we believe is due to our model of diffraction breaking down at high particle concentrations when holograms from objects further from the sensor begin to interfere with objects closer to the sensor. While this could be ameliorated with a thinner sample or improved reconstruction algorithms, we believe the 2 mm sample height used here provides accurate reconstruction over the most relevant concentration range. Of particular importance, the LFI system is sensitive to changes in RBC and WBC concentrations around the critical ranges of microscopic hematuria and pyuria in non-centrifuged urine, highlighted by the red vertical dashed lines in Fig. \ref{fig:McKay_LFI_Figure_003}(g)-(h). These results demonstrate that LFI is a promising technique for measuring hematuria and pyuria with proper calibration to account for under-estimation at the higher concentrations. This capability would provide utility beyond the use case of urinary tract infection screening, as hematuria and pyuria also occur in the setting of acute kidney injury, stones, and malignancy of the genitourinary tract \cite{Mariani1989,Feld1997,Sharp2013}.}
\\

\subsection{E. Coli Concentration} \label{EColi_Concentration_PBS_Results}

\textcolor{black}{Next, we tested if the LFI system could resolve and count individual E. Coli, despite their size being a similar order of magnitude to the wavelength of light. Previous studies applying lensless imaging to bacteria demonstrate very weak scattering that has required the development and application of thin wetting films to improve signal-to-noise ratio (SNR) of the hologram \cite{Poher2011,Allier2010}, which is incompatible with in-line urine analysis. To test the ability of LFI to resolve E. Coli and estimate concentration, we created simple UTI phantoms where E. Coli concentration was varied in PBS spanning above and below the typical UTI concentration threshold of 10$^{5}$ cells/mL. Fig. \ref{fig:McKay_LFI_Figure_004}(a)-(g) shows the holograms from this experiment with increasing concentration from left to right (0, 10$^{3}$, 10$^{4}$, 10$^{5}$, 10$^{6}$, 10$^{7}$, and 10$^{8}$ cells/mL). The corresponding reconstruction of these holograms is shown below each panel of the figure as a log-intensity of the summed projection to enable viewing the results across the wide range of concentrations under study. We see that holograms created by individual E. Coli bacteria and their corresponding reconstructions occur with expected frequencies at concentrations up to 10$^{6}$ cells/mL. Beyond this, we begin to see significant hologram overlap and noise in the corresponding reconstruction. Interestingly, in place of the individual holograms we see instead a textural change appear, that is similar to speckle noise \cite{Goodman2020}.}

\begin{figure*}[h!]
\centerline{\includegraphics[width=\linewidth]{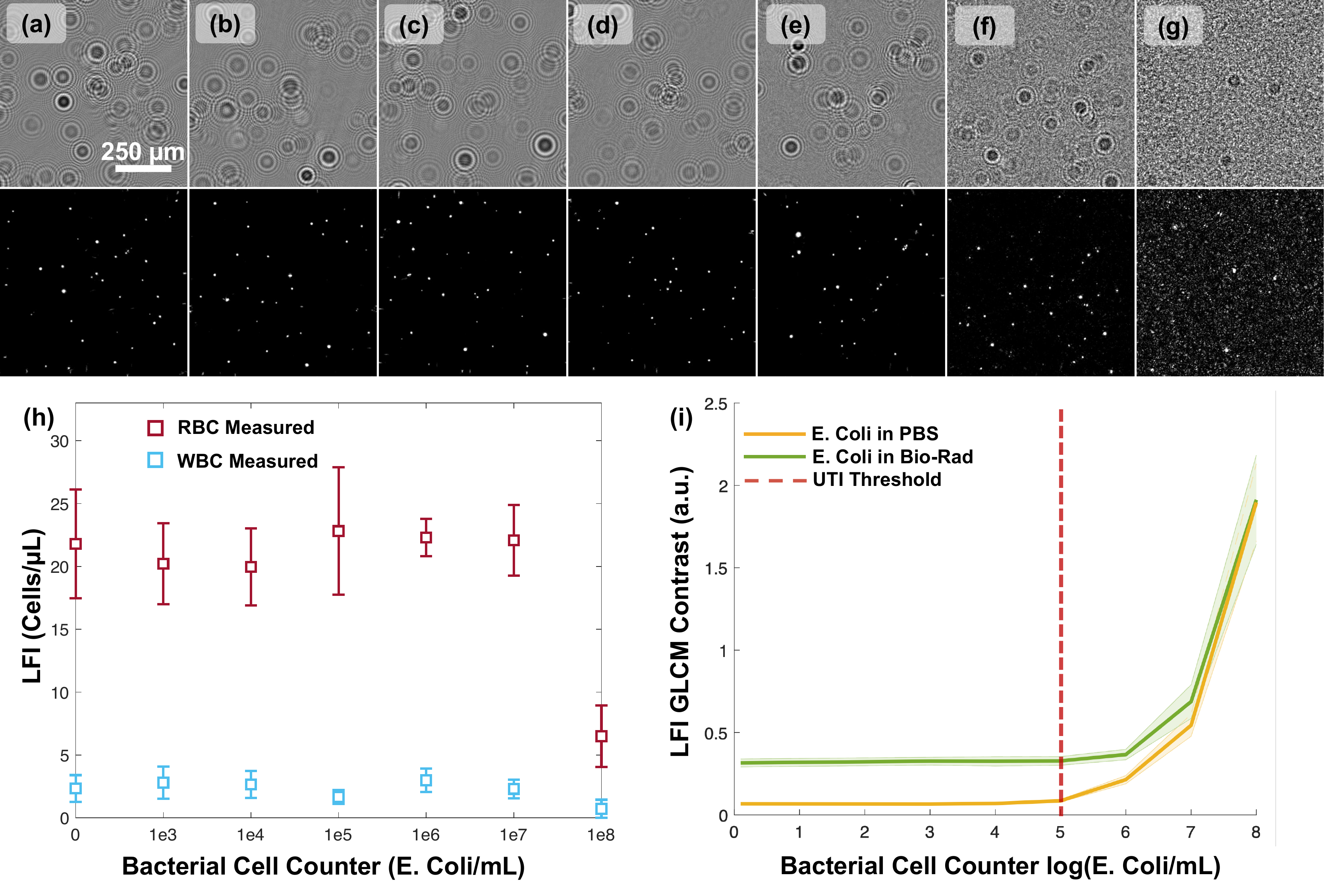}}
\caption{Red blood cell (RBC), white blood cell (WBC), and E. Coli concentration estimation in Bio-Rad urinalysis control. (a)-(g) Holograms (top) and corresponding summed intensity projections of reconstructions (bottom) of increasing concentration of E. Coli (0, 10$^{3}$, 10$^{4}$, 10$^{5}$, 10$^{6}$, 10$^{7}$, and 10$^{8}$ cells/mL, respectively) with constant concentration of RBCs and WBCs. (h) Manual counting of cells yields consistent concentration estimations of RBCs and WBCs until E. Coli concentration of 10$^{8}$ cells/mL. (i) Gray-level co-occurrence matrix contrast vs. E. Coli concentration quantifies textural changes that occur at higher E. Coli Concentrations in the Bio-Rad urinalysis control (green) as compared to PBS (Fig. \ref{fig:McKay_LFI_Figure_004}(i), yellow).}
\label{fig:McKay_LFI_Figure_005}
\end{figure*}

\textcolor{black}{By manually counting the number of E. Coli particles localized in each reconstructed image and dividing by the volume, we generate an estimation of E. Coli concentration from the LFI measurements. When plotting these results against ground truth concentration measurements determined with a Petroff-Hausser bacterial cell counter (Fig. \ref{fig:McKay_LFI_Figure_004}(h)), we see that LFI demonstrates accurate estimation over the most clinically important range of 10$^{3}$ cells/mL to 10$^{6}$ cells/mL. However, at lower concentration, the LFI system over-estimates the number of bacteria, likely from small particles of debris or bubbles in the sample being misclassified as E. Coli. Above 10$^{6}$ cells/mL, the LFI system begins to under-estimate the true E. Coli concentration. This trend is similar to the RBC and WBC concentration estimates of Fig. \ref{fig:McKay_LFI_Figure_003}(g)-(h)), and again occurs at approximately the concentration where individual hologram fringes are no longer resolved and the reconstructions become noisy.}

\textcolor{black}{While individual particle counting fails above the UTI threshold, there is a noticeable change in hologram texture that increases with very high concentrations. We implemented the gray-level co-occurrence matrix (GLCM), a computationally efficient textural analysis tool that quantifies how frequently intensity values occur in pre-defined spatial patterns in the image. We plot the GLCM contrast of the LFI holograms vs. E. Coli concentration in Fig. \ref{fig:McKay_LFI_Figure_004}(i), which demonstrates a clear positive correlation with concentration above the UTI threshold. Thus, using these complementary techniques, manual counting at low concentrations and textural analysis at high concentrations, it may be possible to estimate E. Coli concentration with the LFI system across the full eight orders of magnitude tested.}

\subsection{Measuring cross-talk between E. Coli, RBC, and WBC concentrations} \label{EColi_RBCs_WBCs_Results}

\textcolor{black}{The use of PBS as the solvent for testing E. Coli concentration in Fig. \ref{fig:McKay_LFI_Figure_004} provides an idealized scenario for assessing UTI. However, it ignores the fact that patients often have other large particles such as RBCs and WBCs present in their urine in the setting of UTI. To model possible cross-talk between different particulate concentrations, we next added E. Coli to Bio-Rad urinalysis control diluted in PBS such that the WBC concentration was near the pyuria threshold. We created samples again with concentrations of E. Coli spanning above and below the UTI threshold (0, 10$^{3}$, 10$^{4}$, 10$^{5}$, 10$^{6}$, 10$^{7}$, and 10$^{8}$ cells/mL), and imaged them with the LFI system. Fig. \ref{fig:McKay_LFI_Figure_005} summarizes these results, with (a)-(g) showing the hologram (top) and corresponding reconstruction (bottom) with increasing E. Coli concentration from left to right.}

\textcolor{black}{The SNR from blood cell holograms is much stronger than from the weakly scattering E. Coli particles, which is readily apparent both in the holograms and the reconstructions. While we observed a linear increase in the number of individual E. Coli holograms and corresponding reconstructed bacteria when they were in PBS (Fig. \ref{fig:McKay_LFI_Figure_004}), we do not observe a similar trend when the E. Coli are mixed with blood cells. Unfortunately, it appears that individual E. Coli are not resolvable when mixed with blood cells at this concentration. However, despite this failure, we are able to reconstruct and accurately measure the concentration of RBCs and WBCs as E. Coli concentration increases up through 10$^{7}$ cells/mL.}

\textcolor{black}{Finally, though E. Coli reconstruction and concentration estimation fails below the UTI threshold, we do see the same characteristic change in texture in the holograms at and above 10$^{6}$ cells/mL that we observed when PBS was the solvent (Fig. \ref{fig:McKay_LFI_Figure_004}). Thus, when applying the GLCM textural analysis to the data, we observe a similar trend, which is demonstrated in Fig. \ref{fig:McKay_LFI_Figure_005}(i) as the green curve. For comparison, the yellow curve is the same data from Fig. \ref{fig:McKay_LFI_Figure_004}(i), where E. Coli was suspended in PBS. Note that the strong holograms from RBCs and WBCs provide a vertical offset to the GLCM contrast signal at the low concentrations of E. Coli. However, similar to when PBS is the solvent, the GLCM contrast still correlates with higher concentrations of E. Coli, as speckle-like signal prevails over the stronger SNR holograms from the blood cells. Further, at the highest concentration of E. Coli, we observe a marked drop in the number of reconstructed blood cells, which appears to affect the RBC counting more than WBC counting. White blood cells could be less prone to this phenomenon because they are typically larger than RBCs and have more complex sub-cellular features that diffract light.}

\begin{figure}[h!]
\centerline{\includegraphics[width=\columnwidth]{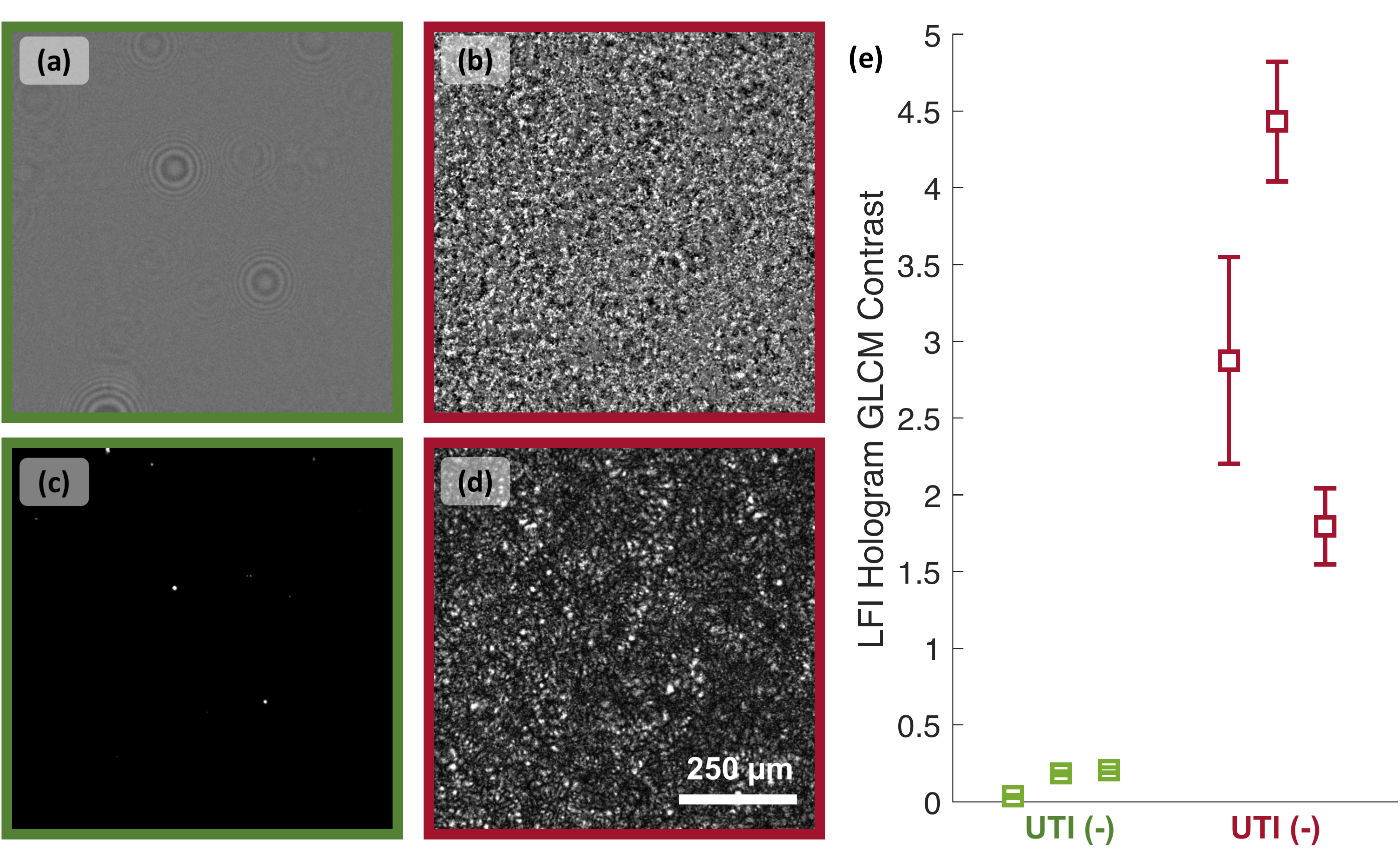}}
\caption{(a) Representative LFI hologram of human urine negative UTI control (green box), and (b) representative LFI hologram of human urine positive UTI (red box), with (c)-(d) corresponding 2D summed intensity projection of the 3D reconstructions. (e) Gray-level co-occurrence matrix contrast analysis applied to six human urine samples show a clear difference between UTI negative controls (green data points) and UTI positive (red data points).}
\label{fig:McKay_LFI_Figure_006}
\end{figure}

\subsection{Human UTI Sample Imaging} \label{UTISamples_Results}
\textcolor{black}{Next, we assessed the difference in LFI measurements between human urine samples with known positive UTI diagnosis (UTI(+)) and known negative UTI controls (UTI(-)). These data are shown in Fig. \ref{fig:McKay_LFI_Figure_006}(a)-(b), where a representative UTI(-) hologram is highlighted in a green box, and a representative UTI(+) hologram is highlighted in a red box. These holograms show a clear difference in signal across the two classes. UTI(+) cases show numerous large, high SNR holograms surrounded by higher spatial frequency texture, which is qualitatively similar to what was observed in our model UTI phantoms with hematuria and pyuria (Fig. \ref{fig:McKay_LFI_Figure_005}(f)-(g)). The 2D summed intensity projections of the corresponding 3D reconstruction from these holograms are shown in Fig. \ref{fig:McKay_LFI_Figure_006}(c)-(d). Overcrowding of the holograms and a breakdown of the sparsity assumption in the UTI(+) cases leads to noisy reconstructions similar to reconstructions observed in Fig. \ref{fig:McKay_LFI_Figure_005}(f)-(g) Next, we applied the GLCM contrast analaysis to five different regions of interest in each of the samples collected. The results are shown in Fig. \ref{fig:McKay_LFI_Figure_006}(e), demonstrating a clear, quantitative textural difference between the UTI(-) cases and the UTI(+) cases. Though these are promising results, the exact sensitivity of LFI to UTI diagnosis is not discernible from this preliminary data because the UTI(+) appear to be at concentrations much higher than the 10$^{5}$ cells/mL threshold. Further, though the UTI(-) cases do show fewer, weaker SNR holograms and less texture, there are still particles present in the sample. These are likely a complex, heterogenous mixture of urothelial cell debris, crystals, sperm, and mucous, and further work must be done to calibrate LFI imaging to the variability present in human urine. It will be interesting to study how phenomena such as hydration, pH, temperature, and time of day affect the urine composition and resulting LFI signal. Thus, we observe promising preliminary results in the ability of LFI to distinguish between UTI(+) and UTI(-) samples, though significant work remains to fully understand the complexity of urine composition and correlation to LFI data in health and disease.}

\section{Conclusion}
\textcolor{black}{Lens free imaging provides a compact, low-cost method of assessing large volumes of weakly scattering material. These strengths make it a natural fit for bedside, point-of-care urinary tract infection screening for hematuria, pyuria, and bacteriuria. In this manuscript, we demonstrate an LFI system that resolves and estimates the concentration red blood cells, white blood cells, and bacteria over clinically important ranges using a 3D sparse phase recovery reconstruction algorithm and textural analysis. Further, we show that LFI holograms show qualitative differences between human urine with positive UTI diagnoses and negative controls. These results demonstrate that LFI is a promising technology for urine tract infection screening. In future work, we aim to implement this technology within a flow cell, directly in-line with an in-dwelling catheter to enable urinalysis screening at the bedside in real-time. Further, we aim to study variability in urine composition and correlates to LFI signal in both health and disease. This technology could alleviate issues with handling human waste and enable real-time trend analysis that yields early detection of UTI, kidney injury, and other conditions in the genitourinary tract. Finally, this technology could be readily adapted to provide urine screening in low-resource settings where conventional laboratory equipment is unavailable.}



\section*{Acknowledgment}

\textcolor{black}{We would like to thank Chirag Parikh, M.B.B.S., Ph.D., Wassim Obeid, Ph.D., Pamela Corona Villalobos, M.D., M.S., and Bill Clarke, Ph.D., for their help acquiring patient samples and insight and guidance interpreting results. The authors are co-inventors on a provisional patent application assigned to Johns Hopkins University. They may be entitled to future royalties from intellectual property related to the technologies described in this article.}

\bibliographystyle{bib_sobraep}
\bibliography{LFI_Urinalysis_Bibliography}
\end{document}